\documentclass[twocolumn,aps,showpacs,floatfix,prc]{revtex4}
\usepackage[dvips]{epsfig}
\begin{document}
\title{$J/\psi$ suppression in the threshold model at RHIC   and LHC   energy }
\author{A. K. Chaudhuri}
\email[E-mail:]{akc@veccal.ernet.in}
\affiliation{Variable Energy Cyclotron Centre,\\ 1/AF, Bidhan Nagar,
Kolkata 700~064, India}

\begin{abstract}
In the QGP based threshold model  \cite{Blaizot:2000ev,Blaizot:1996nq}, in addition to the normal nuclear absorption, $J/\psi$'s are subjected to an 'anomalous' suppression such that above a threshold density $n_{J/\psi}$, all the $J/\psi$'s are melted.
In the threshold model we have analysed
the recent PHENIX data on the centrality dependence of $J/\psi$ suppression in Au+Au collisions at RHIC. Feedback from decay of the state $\chi$ is accounted for. 
  $J/\psi$'s are anomalously suppressed above a threshold density, $n_{J/\psi}=3.57\pm 0.17$ $fm^{-2}$. Threshold density for anomalous suppression of  the state $\chi$ is uncertain to a large extent, $n_\chi=0.32 \pm 0.32$ $fm^{-2}$.  
The fraction $F$  
of the state $\chi$ can not be determined unambiguously, depending on the nuclear absorption, it can vary from 20\% to 40\%.   
We have also predicted for the suppression in Pb+Pb collisions at LHC energy. In central Pb+Pb collisions, $J/\psi$'s are suppressed by a factor of 3-4.   Suppression pattern is rather similar to that in Au+Au collisions, if not slighty less in central collisions.
Using the PHENIX data on the participant number dependence of the Bjorken energy density, we have also estimated the QGP formation time. For critical temperature $T_c$=192  MeV, estimated QGP formation time ranges between 0.07-0.09 fm/c.
\end{abstract}  
\pacs{PACS numbers: 25.75.-q, 25.75.Dw}

\maketitle
 
\section{introduction} 

 In  relativistic  heavy  ion  collisions $J/\psi$ suppression has
been recognized as an important tool  to  identify  the  possible
phase transition to quark-gluon plasma. Because of the large mass
of  the  charm  quarks,  $c\bar{c}$ pairs are produced on a short
time scale. Their tight binding also makes them immune  to  final
state interactions. Their evolution probes the state of matter in
the  early  stage  of the collisions. Matsui and Satz  \cite{Matsui:1986dk}, 
predicted that in presence of quark-gluon plasma  (QGP),  binding
of a $c\bar{c}$  pair  into  a  $J/\psi$  meson will be hindered,
leading to the  so  called  $J/\psi$  suppression  in  heavy  ion
collisions  \cite{Matsui:1986dk} .  Over  the  years,  several  groups have
measured the $J/\psi$ yield in heavy ion collisions (for a review
of the data prior to RHIC energy collisions, and the interpretations see Refs.  \cite{vo99,ge99}).
In  brief,  experimental  data do show suppression. However, this
could be attributed to the conventional nuclear absorption,  also
present in $pA$ collisions.

In  recent  Au+Au  collisions  at  RHIC,  one  observe a dramatic
suppression of hadrons with high  momentum,  transverse  to  beam
direction            (high           $p_T$           suppression)
\cite{BRAHMSwhitepaper,PHOBOSwhitepaper,PHENIXwhitepaper,STARwhitepaper}.
This has been interpreted as an evidence for the  creation  of  high
density,  color  opaque  medium  of  deconfined quarks and gluons
\cite{QGP3jetqu}. It is expected that high density, color opaque medium
will leave its imprint on $J/\psi$ production. At RHIC energy, it has been
argued that rather than suppression, charmoniums will be enhanced
\cite{ Thews:2000rj,Braun-Munzinger:2000px}. 
Due to large initial energy, large number of $c\bar{c}$ pairs will be
produced in initial hard scatterings. Recombination of $c\bar{c}$
can occur enhancing the charmonium production.
Recently, PHENIX  collaboration  have published their measurement of the centrality dependence of
$J/\psi$ suppression in  Au+Au collisions  at  RHIC  energy \cite{Adare:2006ns}. Data are taken
at mid-rapidity ($|y| < .35$) and at forward rapidity ($1.2<y<2.2$). $J/\psi$'s  are more suppressed at forward rapidity than at mid rapidity.
In most central Au+Au collisions, while at mid rapidity, $J/\psi$'s are suppressed by a factor of $\sim 3$, suppression factor is $\sim 6$ at forward rapidity. PHENIX collaboration also measured the centrality dependence of $J/\psi$ suppression in
mid and forward rapidity Cu+Cu collisions \cite{Cianciolo:2006fj}.
Final analysis is not completed yet. Preliminary analysis indicate that,
unlike in Au+Au collisions, $J/\psi$  suppression in Cu+Cu collisions, donot show large dependence on rapidity.
Within the experimental errors, in most central Cu+Cu collisions, 
both at mid rapidity  and forward rapidity. 
$J/\psi$'s are suppressed by a factor of $\sim$ 2.
Both the PHENIX data on
$J/\psi$ production in Au+Au and in Cu+Cu collisions, are not consistent
with models which predict $J/\psi$ enhancement
\cite{ Thews:2000rj,Braun-Munzinger:2000px}.  It was also  seen
that  various  models,  e.g.  comover  model   
\cite{Capella:2000zp}, statistical  coalescence  model    
\cite{Kostyuk:2003kt}  or the kinetic model  \cite{Gorenstein:2000ck,Grandchamp:2003uw},
fail to explain the   (preliminary) PHENIX data  on  the  nuclear
modification   factor   for   $J/\psi$  in  Cu+Cu  and  in  Au+Au
collisions. The data are also not  explained  in  the Glauber model of normal  nuclear
absorption \cite{Vogt:2005ia}.  
The 
QCD based nuclear absorption model 
\cite{Qiu:1998rz,Chaudhuri:2001zx} could explain   the PHENIX data on the centrality dependence of $J/\psi$ suppression in Cu+Cu collisions at RHIC but failed for
Au+Au collisions \cite{Chaudhuri:2006xm}.
  Gunji et al \cite{Gunji:2007uy} analysed the PHENIX data in a hydro+$J/\psi$ model.  The experimental $J/\psi$ suppression pattern in mid-rapidity is well explained by sequential melting of $\chi$, $\psi\prime$ and $J/\psi$ in an dynamically expanding fluid. 
Melting temperatures of $J/\psi$ and $\chi$ are found to be  $T_{J/\psi}=2.02 T_c$,
$T_{\chi}=1.22 T_c$, in agreement with lattice based model calculations \cite{Satz:2006kb}. They estimated the fraction of higher states as $\sim$ 30\% which is also in reasonable agreement with measurements 
 in hadron-hadron collisions \cite{Antoniazzi:1992af,Antoniazzi:1992iv}. 

Recently, in the threshold model \cite{Blaizot:2000ev,Blaizot:1996nq}, we have analysed the PHENIX data on $J/\psi$ suppression \cite{Chaudhuri:2006fe,Chaudhuri:2007qz}.
Blaizot  et  al  \cite{Blaizot:2000ev,Blaizot:1996nq},  proposed the threshold
model  to  explain  the NA50 data \cite{Abreu:2000ni} on  anomalous $J/\psi$ suppression in 
158 AGeV Pb+Pb 
collisions at SPS energy.   
To mimic the onset of deconfining phase transition
above  a  critical  energy  density  and  subsequent  melting  of
$J/\psi$'s, $J/\psi$ suppression was linked with the local transverse 
density.  If  the local transverse density at the point where $J/\psi$ is
formed, exceeds a threshold  value  ($n_{J/\psi}$),  $J/\psi$'s
are melted. 
Our analysis indicate that $J/\psi$ suppression in mid-rapidity Au+Au collisions are well explained in the threshold model for threshold density $n_{J/\psi}=3.73\pm 0.29 fm^{-2}$ \cite{Chaudhuri:2007qz}. Forward rapidity data also require similar threshold density but the threshold density has to be smeared greatly. Since large smearing of the threshold density is contrary to the spirit of the model, it was concluded that while $J/\psi$ suppression  in  Au+Au collisions at mid-rapidity are consistent with the threshold model, that in the forward rapidity are not.  
In \cite{Chaudhuri:2007qz}, with the help of the PHENIX data on centrality dependence of Bjorken energy density, it was argued that the threshold density is related to the  Bjorken energy density times the QGP formation time.  Threshold density $n_{J/\psi}\approx 3.73 fm^{-2}$ then indicate that QGP formation time is $\sim$ 0.06-0.08 fm for critical temperature 192 MeV or less. 
 
The analysis in \cite{Chaudhuri:2006fe,Chaudhuri:2007qz} ignore two important aspect of $J/\psi$ production (i) feedback from higher states $\chi$ and $\psi^\prime$ and (ii) transverse expansion. Threshold model is a static model and transverse expansion can not be incorporated in the model. However, the model can be extended to include the feedback from the higher states.
      In the present paper, we have extended the threshold model to include the feedback from higher states and  analysed the PHENIX data on mid-rapidity $J/\psi$ suppression in Au+Au collisions. We also use the threshold model  to predict for the centrality dependence of $J/\psi$ suppression in Pb+Pb collisions at LHC energy ($\sqrt{s}$=5500 GeV). The model predicts that the centrality dependence of $J/\psi$ suppression  in Pb+Pb collisions is similar to that in Au+Au collisions at RHIC.
 The plan of the paper is as follows: in section II, we briefly describe the threshold model. PHENIX data on the centrality
dependence of $J/\psi$ suppression in Au+Au collisions at mid-rapidity are analysed in section III.   We have used the model to predict the centrality dependence of $J/\psi$ suppression in Pb+Pb collisions at LHC energy (section IV). In section V, we discuss the connection of threshold model with Bjorken energy density times the QGP formation time. Lastly,
  summary and conclusions are drawn in section VI.

\section{The threshold model for $J/\psi$ suppression}

The details of the threshold model could be found in  \cite{Blaizot:2000ev,Blaizot:1996nq}.
It is assumed that fate of a $J/\psi$ depend on the local energy
density, which is proportional to participant density. If the energy density or equivalently, the participant density, exceeds a critical or threshold value, deconfined matter is formed and all the $J/\psi$'s are completely destroyed (anomalous suppression). 
This anomalous suppression is in addition to the
"conventional nuclear absorption".  In conventional nuclear absorption, $J/\psi$ are destroyed due to inelastic $J/\psi$- nucleon collisions.
Transverse expansion of the system is neglected. It is implicitly assumed that $J/\psi$'s are suppressed before the transverse expansion sets in.

In the threshold model, number of $J/\psi$ mesons, produced in a AA collision, at impact parameter ${\bf b}$ can be written as,

\begin{eqnarray} \label{eq1}
\sigma^{J/\psi}_{AA}({\bf b}) = &&\sigma^{J/\psi}_{NN} \int  
 d^2{\bf s}  
 T^{eff}_A({\bf s}) T^{eff}_B({\bf b-s}) \nonumber \\
  &&\times S_{anom}({\bf b,s}),
\end{eqnarray}
  
\noindent  where $T^{eff}(b)$ is the effective nuclear thickness,

\begin{equation} \label{eq2}
T^{eff}({\bf  b})=\int_{-\infty}^{\infty}  dz  \rho({\bf   b},z)
exp(-\sigma_{abs}  \int_z^{\infty} dz\prime \rho({\bf
b},z\prime)),
\end{equation}

\noindent $\sigma_{abs}$ being the $J/\psi$-Nucleon absorption cross-section.  
$S_{anom}({\bf   b,s})$  in  Eq.\ref{eq1}  is  the  anomalous
suppression factor introduced by Blaizot {\em et al.}  \cite{Blaizot:2000ev,Blaizot:1996nq}. Assuming that  all  the
$J/\psi$'s  get suppressed above a threshold density ($n_{J/\psi}$), the
anomalous suppression can be written as,

\begin{equation}  \label{eq3}
S_{anom}({\bf b,s}) =\Theta (n_{J/\psi} - n({\bf b,s})) 
\end{equation}

It was observed \cite{Blaizot:2000ev} that
by smearing the threshold density by a small amount, one can obtain better fit to the data, but at the expense of an additional parameter ($\lambda$),

\begin{equation}  \label{eq4}
S_{anom}({\bf b,s}) =\frac{1}{2} \left[ 1-\tanh \lambda (n({\bf b,s})-n_{J/\psi}) \right ] 
\end{equation}

Critical ingredient of the threshold model is the "local" transverse density. At impact parameter ${\bf b}$ and at the transverse position ${\bf s}$, local transverse density it can be obtained as,

\begin{eqnarray} \label{eq5}
n({\bf b,s})=&&T_A({\bf s})[1-exp(-\sigma_{NN} T_B({\bf s}-{\bf b}))] \nonumber \\
&&+T_B({\bf b}-{\bf s})[1-exp(-\sigma_{NN} T_A({\bf s}))]
\end{eqnarray}
 
\noindent where $\sigma_{NN}$ is the nucleon-nucleon inelastic cross-section. At RHIC energy we use $\sigma_{NN}$=44 mb.
As mentioned previously, the model give excellent description of the PHENIX data on participant number dependence of $J/\psi$ suppression with (i)$\sigma_{abs}\approx$4.39mb and (ii) threshold density $n_{J/\psi}\approx$3.73 $fm^{-2}$ \cite{Chaudhuri:2007qz}.  

The model can be easily extended to include the feedback from higher states $\chi$ and $\psi\prime$. 
Approximately 30\% of $J/\psi$'s are from decay of the state $\chi$ and 10\% from the state $\psi\prime$ \cite{Antoniazzi:1992af,Antoniazzi:1992iv}.  
Lattice model calculations indicate that dissociation temperature for $\chi$ and $\psi\prime$ are similar.   $T_d^{\chi} \approx 1.2 T_c$ and $T_d^{\psi\prime} \approx 1.1 T_c$ \cite{Satz:2006kb}. 
$J/\psi$'s can survive much higher temperature $T_{J/\psi}\approx 2.1 T_c$.
We ignore the small difference between dissociation temperatures of $\chi$ and $\psi^\prime$  and assume that fraction $F$ of $J/\psi$ are from decay of the higher state
$\chi$.  Both the state $J/\psi$ and $\chi$ can be anomalously suppressed. And like the state $J/\psi$, 
the state $\chi$ could also be suppressed  due to inelastic collisions with the nucleons.  $\chi$-nucleon absorption cross-section ($\sigma_{abs}^\chi$) need not be same as that for $J/\psi$-nucleon absorption cross-section.  
If fraction $F$ of the observed $J/\psi$ are from decay of the higher state $\chi$, the $J/\psi$ production at 
impact parameter {\bf b} can be written as,

\begin{widetext} 
\begin{eqnarray} \label{eq6}
\sigma^{J/\psi}_{AA}({\bf b}) =  
\sigma^{J/\psi}_{NN} \left[(1-F) \int
 d^2{\bf s}  
 T^{eff}_A({\bf s}) T^{eff}_B({\bf b-s})  
  \times S_{anom}({\bf b,s})
    + F \int  
  d^2{\bf s}  
 {T^\prime}^{eff}_A({\bf s}) {T^\prime}^{eff}_B({\bf b-s}) 
  \times S^{\chi}_{anom} ({\bf b,s}) 
\right]
\end{eqnarray}
\end{widetext}
 
\noindent where the first term is as usual for the direct $J/\psi$ production. The 2nd term accounts for the feedback from the state $\chi$. ${T^\prime}^{eff}$ is the effective nuclear thickness for the  state $\chi$,
 
\begin{equation} \label{eq7}
{T^\prime}^{eff}({\bf  b})=\int_{-\infty}^{\infty}  dz  \rho({\bf   b},z)
exp(-\sigma^{X}_{abs}  \int_z^{\infty} dz\prime \rho({\bf
b},z\prime)),
\end{equation}
 
\noindent and,
$S^\chi_{anom}$ is the anomalous suppression factor for the state $\chi$,

\begin{equation}  \label{eq8a}
S^{\chi}_{anom}({\bf b,s}) =\Theta(n_\chi - n({\bf b,s}))
\end{equation}

\noindent with threshold density $n_\chi$. The $\Theta$ function can again be smeared at the expense of an additional 
parameter $\lambda^\chi$,

\begin{equation}  \label{eq8}
S^{\chi}_{anom}({\bf b,s}) =\frac{1}{2} \left[ 1-\tanh \lambda^\chi (n({\bf b,s})-n_\chi) \right ] 
\end{equation}

\section{$J/\psi$ suppression in Au+Au collisions}

Threshold model as presented here has quite a few number of  parameters, $n_{J/\psi}$ and $n_\chi$, the threshold densities for anomalous suppression of $J/\psi$ and $\chi$,the fraction $F$ of the state $\chi$,  $\sigma_{abs}$ and $\sigma^{\chi}_{abs}$ the $J/\psi$-nucleon and $\chi$-nucleon absorption cross sections. Two more parameters, $\lambda$ and $\lambda^\chi$ are added if  smeared Theta functions are used for the anomalous suppression of $J/\psi$ and $\chi$.   
To reduce the number of parameters, we make the  simplifying assumptions that 
$J/\psi$-nucleon or $\chi$-nucleon absorption cross section is proportional to square of the radius of the $J/\psi$ or the $\chi$ meson,

\begin{equation}\label{eq9}
\sigma^\chi_{abs} =\left(\frac{r_\chi}{r_{J/\psi}}\right)^2 \sigma_{abs}
\end{equation}

Radius of different charmonium states has been calculated,  $r_{J/\psi}$=0.25 fm and $r_{\chi}$=0.36 fm \cite{Satz:2006kb}. The  parameter
  $\sigma^\chi_{abs}$ is then expressed in terms of $\sigma_{abs}$, $\sigma^\chi_{abs}=2.07\sigma_{abs}$.

\begin{figure}
\includegraphics[bb=44 284 524 769
 ,width=0.9\linewidth,clip]{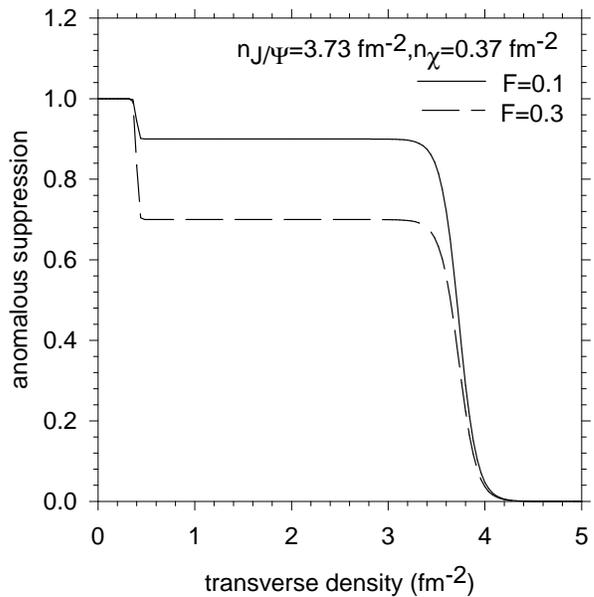}
\caption{Anomalous suppression as a function of transverse density, for threshold density $n_{J/\psi}$=3.73 $fm^{-2}$ and
$n_\chi$=0.37 $fm^{-2}$.  The solid and dashed line corresponds to $F$=0.1 and F=0.3 respectively.  }
 \label{F1}
\end{figure}

We  also fix the smearing of the Theta functions to  fixed values,  
$\lambda=0.05n_{J/\psi}$,
$\lambda_{\chi}=0.05n_\chi$. Since dissociation temperature of the state $\chi$ is approximately half of the  dissociation temperature of the state $J/\psi$ \cite{Satz:2006kb} and in the threshold model, threshold densities are proportional to energy density,  it is expected that the threshold
density for the state $\chi$ will be approximately one-tenth of the 
threshold density for the state $J/\psi$.
In Fig.\ref{F1} we have shown the anomalous suppression $S=(1-F)S_{anom}+F S^\chi_{anom}$ for a representative threshold density $n_{J/\psi}$=3.73 $fm^{-2}$
and $n_{\chi}$=0.37 $fm^{-2}$. The smoothened theta functions retain the basic feature of the threshold model, i.e. sudden onset of anomalous suppression of $J/\psi$ and $\chi$ above  threshold density $\sim$3.73 $fm^{-2}$ and $\sim$ 0.37 $fm^{-2}$ respectively.

\begin{figure}
\includegraphics[bb=19 255 530 741
 ,width=0.9\linewidth,clip]{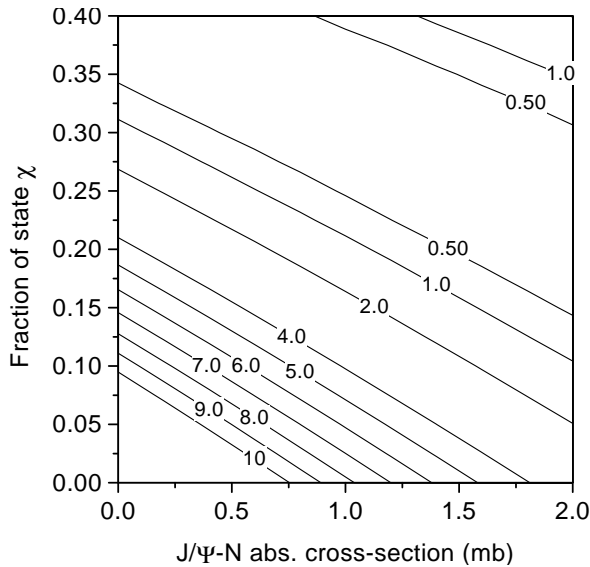}
\caption{contours of constant$\chi^2/N$ in 
$\sigma_{abs}-F$ plane.  The threshold densities are is fixed at $n_{J/\psi}$=3.6 $fm^{-2}$ and $n_\chi$=0.36 $fm^{-2}$. One observe that equivalent $\chi^2/N$ is obtained by increasing F and decreasing $\sigma_{abs}$ or the vice-versa.}
 \label{F2}
\end{figure}

With the simplifying assumptions, number of undetermined parameters in the threshold model is four, $n_{J/\psi}$,
$n_\chi$, $F$ and $\sigma_{abs}$.
However, the parameters   $\sigma_{abs}$ and $F$  can not be determined  simultaneously from a fit to the PHENIX data on the centrality dependence of $J/\psi$ in Au+Au collisions.  In the threshold model $F$ and $\sigma_{abs}$ are independent and one can fit the data either with small F and large $\sigma_{abs}$ or the vice versa. 
This can be seen from   the contour plot of $\chi^2$ shown in Fig.\ref{F2}. As usual,  $\chi^2$ is defined as,

\begin{equation}
\chi^2/N=\frac{1}{N}\Sigma_{i=1}^{i=N} \frac{\left(ex(i)-th(i)\right)^2}{err(i)^2}
\label{eq10}
\end{equation}
\begin{figure}
\includegraphics[bb=71 322 527 793
 ,width=0.9\linewidth,clip]{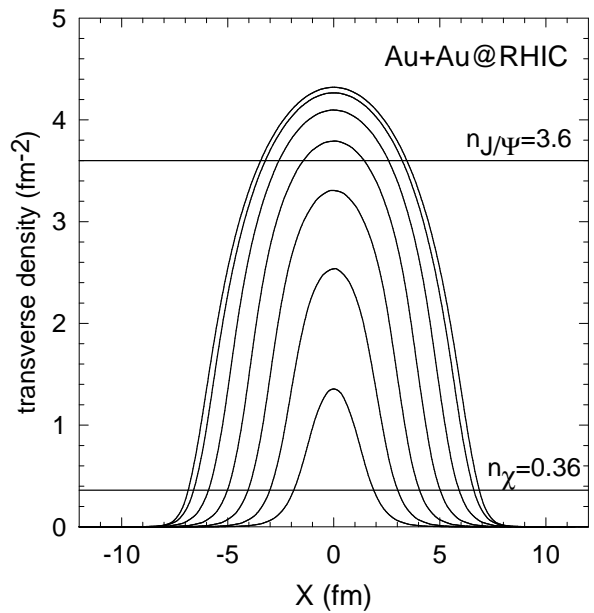}
\caption{Transverse density in Au+Au collisions for various values of the impact parameter b=0,2,4,6..(fm) (top to bottom). The origin is at a distance $d=b/2$ from the center of the nucleus A.
Two solid lines indicate the threshold densities for the state 
$J/\psi$ and the state $\chi$ respectively.}
 \label{F3}
\end{figure}

\noindent where $ex(i)$ and $err(i)$ are PHENIX measurements \cite{Adare:2006ns} for
$J/\psi$ suppression as function of collision centrality and $th(i)$ is the threshold model prediction. A perfect fit is obtained when $\chi^2 \leq 1$. In Fig.\ref{F2}  contours of constant $\chi^2/N$ in $F-\sigma_{abs}$ plane, for  fixed values of the threshold densities, $n_{J/\psi}$=3.6 $fm^{-2}$ and $n_\chi$=0.36 $fm^{-2}$ is shown.
For $\sigma_{abs}$ varying between 0-2 mb, perfect fit to the data could be obtained with $F$ varying between 
0.4-0.1. Equivalent fit ($\chi^2/N\approx 1$) could be obtained either with $F\approx$ 0.3, $\sigma_{abs}\approx$0 mb or with $F\approx$0.1 and $\sigma_{abs}\approx$ 2 mb. The reason is understood also.  
 In the threshold model, suppression depends on the transverse density.  
In Fig.\ref{F3}, we have shown the transverse density $n_p(b,s)$ in Au+Au collisions at impact parameter, b=0,2,4,6,8,10 and 12 fm.  Threshold density for anomalous suppression of the state $\chi$ being small    anomalous suppression can occur in peripheral collisions also.  In contrast, anomalous $J/\psi$ suppression can occur only in sufficiently central ($b < 6 fm$) 
collisions.
Unlike for direct $J/\psi$, anomalous suppression for the state $\chi$ occur right from peripheral to most central collisions. The suppression is rather like nuclear absorption in the Glauber model. In the Glauber model, in all centrality ranges of collisions,
$J/\psi$'s are suppressed.   Naturally, anomalous suppression of the state $\chi$ can approximately compensate for the nuclear absorption of $J/\psi$ (and $\chi$). 
 
\begin{figure}
\includegraphics[bb=38 285 532 769
 ,width=0.9\linewidth,clip]{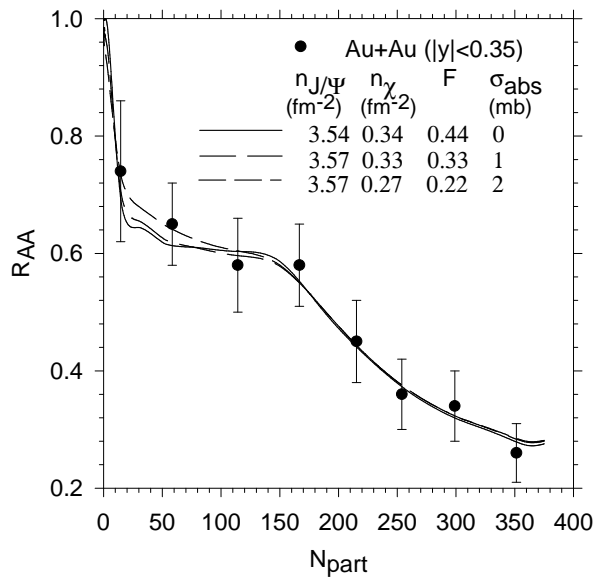}
\caption{Solid circles are PHENIX data \cite{Adare:2006ns} on the centrality dependence of $J/\psi$ suppression in Au+Au collisions . The lines are threshold model fit to data with parameter sets 
depicted in the figure. Note that the parameter $\sigma_{abs}^{J/\psi}$ was kept fixed during the fit.}
 \label{F4}
\end{figure}


Understanding that both $\sigma_{abs}$ and $F$ can not be determined simultaneously from the PHENIX data, for a fixed $\sigma_{abs}$ in the range 0-2 mb, we fit the data by varying $n_{J/\psi}$, $n_\chi$ and $F$.
In Fig.\ref{F4}, fit obtained to the data for $\sigma_{abs}$=0,1 and 2 mb are shown. Parametric values of $n_{J/\psi}$, $n_\chi$ and $F$ are also depicted.
 The threshold model with   higher state feedback correctly reproduces
the centrality dependence of $J/\psi$ suppression in PHENIX measurements. 
We note that within the experimental errors, $J/\psi$ suppression in very peripheral as well as in central collisions is well reproduced even with $\sigma_{abs}$=0 mb.   However, the data require largest contribution from the higher state, $F$=0.44, when the nuclear absorption is neglected.
The model also reproduces the sudden change in the slope  above the participant number $N_{part}\sim 150$.  
It is interesting  to note that different parameter sets predict  nearly identical suppression above $N_{part}\approx $150 . Rather the parameter sets are somewhat distinguished in collisions with $N_{part} <$150.  Apparently, less central collisions are more sensitive to the details of the model rather than very central collisions.
 
\begin{figure}
\includegraphics[bb=38 285 532 769
 ,width=0.9\linewidth,clip]{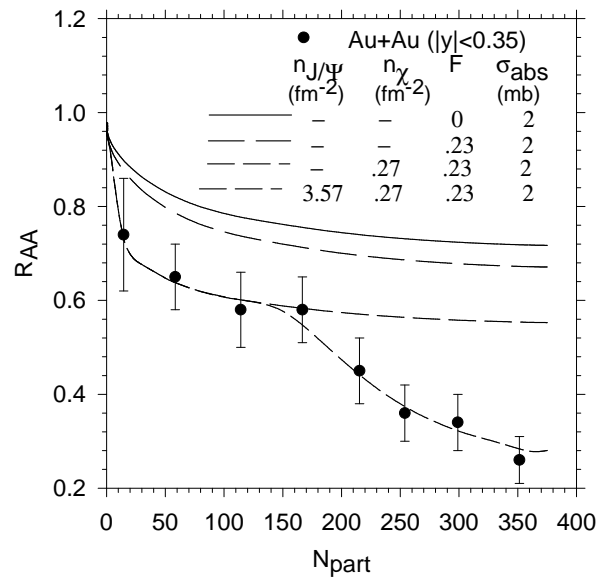}
\caption{Contribution of various components of the threshold model to $J/\psi$  suppression in Au+Au collisions at RHIC are shown. The solid circles are PHENIX data \cite{Adare:2006ns} on the centrality dependence of $J/\psi$ suppression in Au+Au collisions.  }
 \label{F5}
\end{figure}

To understand better the suppression due to various components of the model,
in Fig.\ref{F5}, we have separately shown the suppression due to 
(i) nuclear absorption of $J/\psi$ (the solid line), (ii) nuclear absorption of
$J/\psi$+$\chi$ (the dashed line), (iii) nuclear absorption of
$J/\psi$+$\chi$ and the anomalous suppression of $\chi$ (the medium dashed line) and 
(iv) nuclear absorption of
$J/\psi$+$\chi$ and the anomalous suppression of $J/\psi$ and $\chi$ (the short dashed line). Without any feedback from the state $\chi$, if $J/\psi$'s are suppressed due to nuclear effects only
($\sigma_{abs}$=2 mb), PHENIX data in peripheral collisions or in central collisions are not explained. 
   Suppression is marginally increased ($\chi$-nucleon absorption cross-section is assumed to be twice the $J/\psi$-nucleon absorption cross-section) when 23\% feedback from the state $\chi$ is included.   
Suppression is further increased when the state $\chi$ is allowed to undergo anomalous suppression ($n_\chi$=0.27 $fm^{-2}$) in addition to the nuclear absorption of $\chi$ and $J/\psi$.
  It can be seen   that $\chi$'s are anomalously suppressed nearly in all the centrality ranges of collisions. Anomalous suppression of $\chi$ alone can explain the PHENIX data in peripheral and mid-central collisions ($N_{part} \leq 150$). Data beyond $N_{part}=150$ is underpredicted.
Only when $J/\psi$'s are also suppressed anomalously, 
PHENIX data throughout the centrality range is explained.
For $n_{J/\psi}$=3.57 $fm^{-2}$, direct $J/\psi$'s are anomalously suppressed only in collisions with $N_{part} \geq 150$. Since, anomalous suppression is related to melting of $J/\psi$ in deconfined medium,
the analysis strongly suggests that  sufficiently hot ($T \geq 2.1 T_c$) deconfined medium is created 
in central and mid-central ($N_{part} \geq 150$) Au+Au collisions.

\begin{figure}
\includegraphics[bb= 28 32 528 770
 ,width=0.9\linewidth,clip]{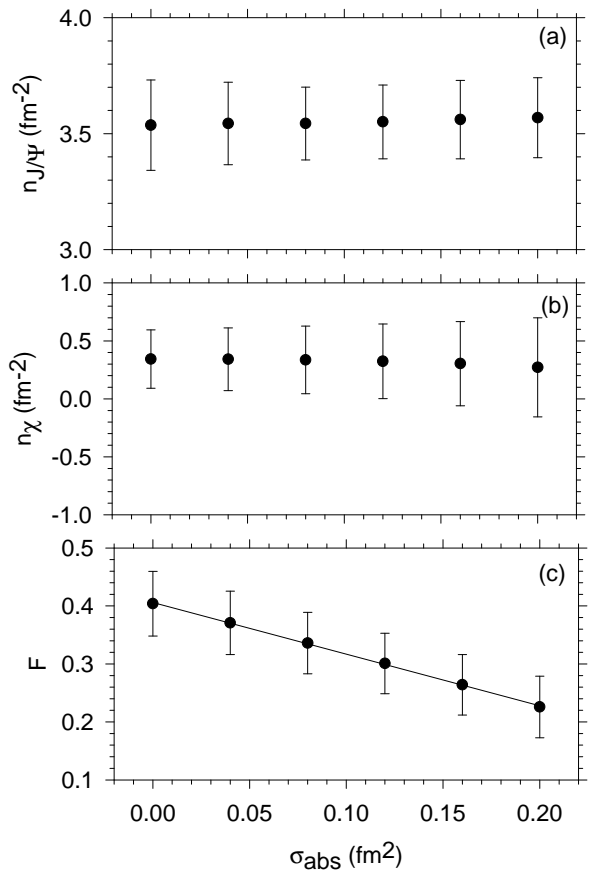}
\caption{In panels (a), (b) and (c) best fitted values for the threshold density for direct $J/\psi$, threshold density for the state $\chi$ and the fraction $F$ of the state $\chi$ are shown for various values of $J/\psi$-nucleon absorption cross section.
The solid line in the panel (c) corresponds to the empirical relation Eq.\ref{eq12}.}  
 \label{F6}
\end{figure}

Variation of the fitted parameters ($n_{J/\psi}$,$n_\chi$ and F) with $\sigma_{abs}$ is interesting.
In panel (a), (b) and (c) of Fig.\ref{F6} we have plotted the best fitted threshold densities $n_{J/\psi}$, $n_\chi$ and the fraction $F$ of the state $\chi$. It may be noted that the uncertainty in the parametric values are large. The quality of the PHENIX data is not good, data points are few and   the error bars are also large.
The uncertainty in the parametric values will reduce with better quality data. It is interesting to note that for $\sigma_{abs}$=0-2 mb, the
 threshold densities for the anomalous suppression of direct $J/\psi$  and also of the state $\chi$ is approximately a  constant. 
The direct $J/\psi$ are anomalously suppressed beyond a threshold density $n_{J/\psi}=3.57\pm 0.17$ $fm^{-2}$. Within the uncertainty, the value is consistent with  $n_{J/\psi}=3.73\pm 0.29 fm^{-2}$, extracted neglecting the feedback from the state $\chi$ \cite{Chaudhuri:2007qz}.   Uncertainty is the  
threshold density for the   state
$\chi$ is quite large, $\chi$ states are anomalously suppressed beyond $n_\chi=0.32  \pm 0.32 fm^{-2}$. 
Since in the threshold model threshold density is assumed to be proportional to the energy density above which the charmonium states get dissociated in QGP, if $T_{J/\psi}$ and $T_\chi$ are the dissociation temperatures then  
$(\frac{T_{J/\psi}}{T_\chi})^4 \approx$
$\frac{n_{J/\psi}}{n_\chi}=11.16 \pm 11.69$. As mentioned earlier, threshold temperatures of $J/\psi$ and $\chi$ are calculated in the lattice based models, $T_{J/\psi} \approx 2.12 T_c$ and $T_\chi\approx 1.2 T_c$ \cite{Satz:2006kb}. Within the large uncertainty, the ratio of the threshold temperatures extracted from the RHIC data on $J/\psi$ suppression agree with the lattice calculation.

While the threshold densities  $n_{J/\psi}$ and $n_\chi$ are found to be nearly independent of $\sigma_{abs}$, the fraction $F$ is not. $F$ linearly decreases with increasing $\sigma_{abs}$. 
As discussed earlier, in the threshold model $\sigma_{abs}$ and $F$ are inter-related and one can compensate for the other.
Depending on the value of $\sigma_{abs}$, the PHENIX data can be well described with $F=0.3\pm 0.1$, the maximum (F=0.4) and minimum (F=0.2) values corresponding to $\sigma_{abs}$=0 and 2  mb respectively.  We also observe that an empirical relation,

\begin{equation}
F=0.406-0.89\sigma_{abs} (fm^2),
\label{eq12}
\end{equation}

\noindent exist between best fitted $F$ and $\sigma_{abs}$.  
It is also interesting to note that the maximum value of $F\approx$0.4  
obtained presently ignoring the nuclear absorption agree 
with the fraction $F$ measured in hadron-hadron collisions \cite{Antoniazzi:1992af,Antoniazzi:1992iv}, where the nuclear absorption is absent by default. 
 
\begin{figure}
\includegraphics[bb=40 287 533 769
 ,width=0.9\linewidth,clip]{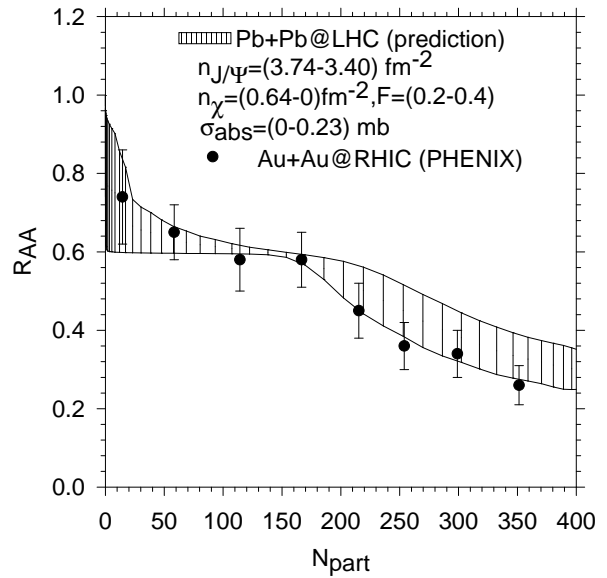}
\caption{
Threshold model predictions for the centrality dependence of $J/\psi$ suppression in Pb+Pb collisions at LHC energy ($\sqrt{s}$=5500 GeV). The shaded region marks the uncertainty in the parametric values of $n_{J/\psi}$, $n_\chi$ and $F$. 
For comparison PHENIX data \cite{Adare:2006ns} on the centrality dependence of $J/\psi$ suppression in Au+Au collisions at RHIC energy are also shown.}
 \label{F7}
\end{figure}

\section{$J/\psi$ suppression in Pb+Pb collisions at LHC}

\begin{figure}
\includegraphics[bb=56 321 531 793
 ,width=0.9\linewidth,clip]{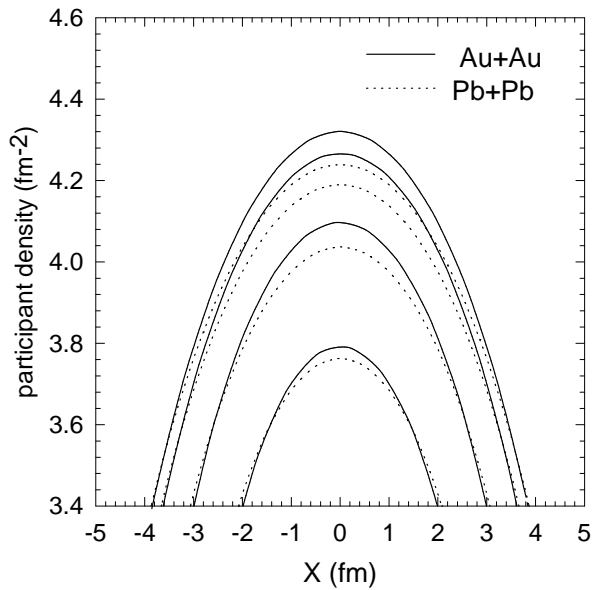}
\caption{The solid and the dotted line show the transverse density in b=0,2,4 and 6 fm Au+Au collisions at RHIC and in Pb+Pb collisions at LHC energy. 
 The origin is at a distance $d=b/2$ from the centre of the nucleus A.
}
 \label{F8}
\end{figure}

At the Large Hadron Collider (LHC), it is planned to collide Pb ions at centre of mass energy  $\sqrt{s}$=5500GeV. Signals of the deconfinement phase transition are expected to  be better defined at LHC than at RHIC. One wonders whether in LHC energy collisions, $J/\psi$ will be more suppressed than at RHIC energy?
The threshold model with the parameters fixed from the
analysis of RHIC data   can be used to predict the suppression in Pb+Pb collisions at LHC.  
  RHIC data on $J/\psi$ suppression are explained with,  $n_{J/\psi}=3.57\pm 0.17 fm^{-2}$ and $n_\chi=0.32 \pm 0.32 fm^{-2}$, $\sigma_{abs}=1\pm 1 mb$ and $F=0.3\pm 0.1$ .    $n_{J/\psi}$ and $n_\chi$  are independent of energy and their value remain unchanged 
  from RHIC to LHC energy.  
However,
the parameter $\sigma_{abs}$  could change from RHIC to LHC energy collisions.  
In SPS energy ($\sqrt{s}$=17 GeV) $\sigma_{abs}\approx$ 4.4 mb \cite{Cortese:2003iz}. It is reduced in RHIC energy ($\sqrt{s}$=200 GeV) collisions, $\sigma_{abs}=1\pm 1 mb$.
 At LHC energy ($\sqrt{s}$=5500 GeV) $\sigma_{abs}$ could be even lower.  However, as shown in Fig.\ref{F2} and Fig.\ref{F6}c, in the threshold model, the parameters $F$ and $\sigma_{abs}$ are not independent, an empirical relation (Eq.\ref{eq12}) exist between $F$ and $\sigma_{abs}$. We assume that at LHC energy such an relation exist between $\sigma_{abs}$ and $F$ and given the fraction $F$, $\sigma_{abs}$ can be determined using  Eq.\ref{eq12}.
   Shaded region 
in Fig.\ref{F7}, show the predicted centrality dependence of $J/\psi$ suppression with $n_{J/\psi}=3.57\pm 0.17$ $fm^{-2}$, $n_\chi=0.32\pm 0.32$ $fm^{2}$ and $F=0.3\pm0.1$.
At LHC energy, for the  nucleon-nucleon inelastic cross-section $\sigma_{NN}$  we have used the value $\sigma_{NN}$=70 mb.
In central collisions, $J/\psi$'s are suppressed by a factor of 3-4.
For comparison, in Fig.\ref{F7}  we have also shown the PHENIX data, which as shown earlier are well explained in the threshold model.  
Interestingly, threshold model predicts that   in central Pb+Pb collisions at LHC, $J/\psi$'s are similarly suppressed (if not slightly less) as in Au+Au collisions at RHIC. Reason 
could be understood as follow: in the threshold model, anomalous suppression depends on the (local) transverse density. Transverse  density depends on the nucleon-nucleon inelastic cross- section, and changes from RHIC to LHC energy. In Fig.\ref{F8}, the transverse density in Au+Au collisions at RHIC and in Pb+Pb collisions at LHC is compared. In central and mid central collisions participant density is less in Pb+Pb collisions than in Au+Au collisions. Then
for a threshold density $n_{J/\psi}$, while in Au+Au collisions $J/\psi$'s could be anomalously suppressed say in a impact parameter $b\prime$ collision, they may not be anomalously suppressed  in Pb+Pb collisions. Thus $J/\psi$'s will be less suppressed in LHC energy. Indeed, explicit calculations indicate that under identical condition,  in central collisions anomalous suppression decreases from RHIC to LHC energy. 

\section{QGP formation time} 

QGP formation time is an important parameter in ultra-relativistic heavy ion collisions. For example, in the  Bjorken 
model with longitudinal boost-invariance, the energy density is obtained as,  

\begin{equation}\label{eq13}
\varepsilon_{BJ}=\frac{1}{\tau A_T} \frac{dE_T}{dy}
\end{equation}

\noindent where $\tau$ is the formation time, $A_T$ is the overlap area and $\frac{dE_T}{dy}$ is the transverse energy per unit rapidity.  
An estimate of energy density produced in a collision then depends
on the formation time $\tau$. QGP formation time ($\tau$) is unknown. Due to heavy mass, $J/\psi$'s are produced early in the collisions. Time scale of $J/\psi$ formation $\tau \sim 1/m_{J/\psi} \sim$ 0.06 fm,  
and experimental data on $J/\psi$ suppression can  give information on the QGP formation time. 

\begin{figure}
\includegraphics[bb=51 274 491 720
 ,width=0.99\linewidth,clip]{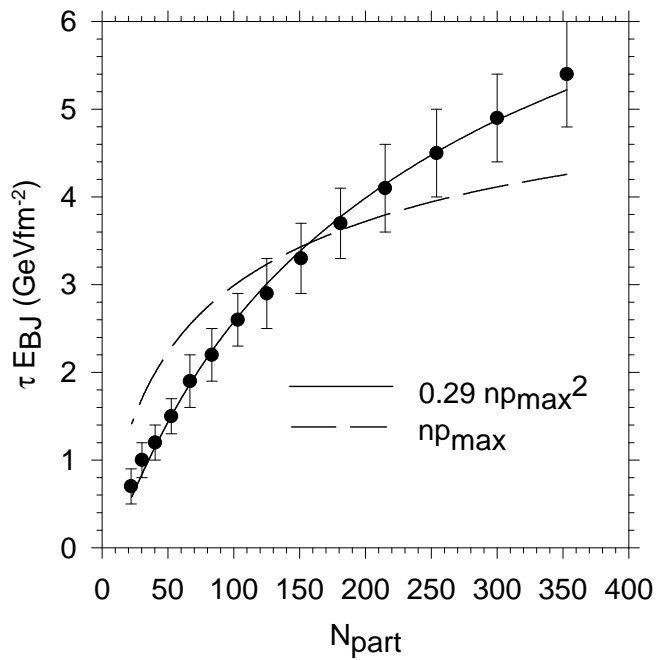} 
\caption{Solid circles are PHENIX data \cite{Adler:2004zn} on the participant number  dependence of Bjorken energy density times formation time ($\tau E_{BJ}$) in Au+Au collisions at $\sqrt{s}$=200 GeV. The dashed line is the participant number dependence of the maximum
transverse density (${n_p}_{max}$) in a Au+Au collision. The solid line is the participant number dependence of $0.29{n^2_p}_{max}$. Empirical relation $\tau \varepsilon_{BJ}=0.29 {n^2_p}_{max}$ correctly reproduces the PHENIX estimate of Bjorken energy density times formation time.}
\label{F9}
\end{figure}

PHENIX collaboration have measured the transverse energy $E_T$
in Au+Au collisions.
Since QGP formation time is not known, they have tabulated the
Bjorken energy density times the formation time as a function of the participant number. In Fig.\ref{F9} we have shown the PHENIX results for the participant number dependence of $\tau \varepsilon_{BJ}$ \cite{Adler:2004zn}. $\tau \varepsilon_{BJ}$ increases with the collision centrality.  

In the threshold model local transverse density (see Eq.\ref{eq5})
is the key ingredient. If the local transverse density exceed the
threshold density $n_{J/\psi}$, $J/\psi$'s are completely suppressed.
It is implicitly assumed that the transverse density is proportional to energy density and in a deconfined matter $J/\psi$'s are melted above a threshold energy density  $\varepsilon_{TH}$.  
As given in Eq.\ref{eq5}, local transverse density is a function of the impact  parameter (${\bf b}$) and the transverse position (${\bf s}$).
In a collision between two identical nuclei at impact parameter ${\bf b}$, maximum transverse density (${n_p}_{max}$) is achieved at the transverse position ${\bf s}={\bf b}/2$. If in a collision  $J/\psi$'s are anomalously suppressed, at the minimum 
${n_p}_{max}$ should exceed $n_{J/\psi}$, ${n_p}_{max} > n_{J/\psi}$.  
In Fig.\ref{F9} the dashed line show the participant number dependence of the maximum transverse density ${n_p}_{max} =n_p({\bf b},{\bf s}={\bf b}/2)$. ${n_p}_{max}$ increases with collision centrality. 
The solid line in Fig.\ref{F9} indicate that empirically, in all the centrality ranges of collisions PHENIX data \cite{Adler:2004zn}
on $\tau \varepsilon_{BJ}$  exhibit a simple relationship with the maximum transverse density ${n_p}_{max}$,

\begin{equation}\label{eq14}
\tau \varepsilon_{BJ} = 0.29 {n^2_p}_{max}
\end{equation}

It is an important observation. We have checked that 
the relation $\tau \varepsilon_{BJ} \propto {n^2_p}_{max}$
is consistent with measurements of $\tau \varepsilon_{BJ}$ in other energies e.g. $\sqrt{s}$=130 GeV and 19.6 GeV also. 
Since ${n_p}_{max} \geq n_{J/\psi}$, empirical relation (Eq.\ref{eq14}) imply, $\tau \varepsilon_{BJ} \leq 0.29 n^2_{J/\psi}$.
If the threshold energy density for $J/\psi$ dissociation is $\varepsilon_{TH}$ then it is necessary that,

\begin{equation}\label{eq15}
0.29 n^2_{J/\psi} \geq \tau \varepsilon_{BJ}  \geq \tau \varepsilon_{TH}
\end{equation}

In otherwords, the threshold density in the threshold model give the upper limit of the threshold energy density for $J/\psi$ dissociation times the QGP formation time.  
Assuming that the $J/\psi$'s are melted above a temperature $T_{TH}=2.1 T_c$, given a critical temperature,
the standard
relation between energy density and temperature,
   
\begin{eqnarray} \label{eq16}
\varepsilon_{TH}=g_{QGP} \frac{\pi^2} {30} T_{TH}^4  
\approx  g_{QGP} \frac{\pi^2} {30} (2.1T_c)^4,
\end{eqnarray}

\noindent  can be used to compute the upper limit of the QGP formation time. 

\begin{table} 
\caption{ Formation time $\tau$ for a choice of critical temperatures for two and three flavored QGP.}
  \begin{tabular}{|c|c|c|} \hline
 &  2-flavored QGP & 3-flavored QGP\\
&  $g_{QGP}$=37 & $g_{QGP}$=47.5\\
$T_c$ (MeV) &  $\tau$(fm) & $\tau$(fm) \\
  \hline
     140.00  &$    0.31  \pm    0.03$ &$ 0.24  \pm    0.02$\\
    150.00  &$    0.24  \pm    0.02$ &$ 0.18  \pm    0.02$\\
    160.00  &$    0.18  \pm    0.02$ &$ 0.14  \pm    0.01$\\
    170.00  &$    0.14  \pm    0.01$ &$ 0.11  \pm    0.01$\\
    180.00  &$    0.11  \pm    0.01$ &$ 0.09  \pm    0.01$\\
    190.00  &$    0.09  \pm    0.01$ &$ 0.07  \pm    0.01$\\
    200.00  &$    0.07  \pm    0.01$ &$ 0.06  \pm    0.01$\\
    210.00  &$    0.06  \pm    0.01$ &$ 0.05  \pm    0.00$\\
    220.00  &$    0.05  \pm    0.00$ &$ 0.04  \pm    0.00$\\
    230.00  &$    0.04  \pm    0.00$ &$ 0.03  \pm    0.00$\\
    240.00  &$    0.04  \pm    0.00$ &$ 0.03  \pm    0.00$\\
    \hline
         \end{tabular} \\
\label{table1}
 \end{table}
      
In table. \ref{table1}, for threshold density $n_{J/\psi}=3.57\pm 0.17$ $fm^{-2}$,
we have tabulated the upper limit of the formation time $\tau$ for a two and three flavored QGP. QGP formation time is low. For critical temperature 
$T_c$=140-200 MeV, it varies from 0.07(0.06)-0.31(0.24) fm.
Critical temperature for the confinement-deconfinement transition has been accurately estimated in recent lattice calculations, $T_c = 192\pm 7\pm4$ MeV \cite{Cheng:2006qk}. As seen from table.\ref{table1}, $T_c \sim 192$ MeV corresponds to formation time $\tau$ ranging between 0.09(0.07) fm for 2(3) flavored QGP. Formation time $\tau$ is rather small compared to initial time $\tau_i$=0.6 fm obtained from hydrodynamic analysis of experimental data in Au+Au collisions  at RHIC \cite{Kolb:2003dz}. However, we note that
the formation time scale 0.09(0.07) fm is rather consistent with Color Glass Condensate (CGC) models \cite{McLerran:2007bn}.  

Before we summarise our results, we would like to mention a deficiency of the model.  
Transverse expansion of the system is neglected, it is assumed that all the $J/\psi$'s are suppressed before the transverse expansion sets in. 
However, hydrodynamics model simulations of Au+Au collisions at RHIC indicate that even the QGP phase undergoes certain amount of transverse expansion.  
  With transverse expansion, transverse density is diluted and $J/\psi$'s which are anomalously suppressed in a static medium may not be suppressed in an expanding medium. 
Thus the threshold density $n_{J/\psi}=3.57\pm 0.17 fm^{-2}$ represent an upper limit to threshold density. The QGP formation time as tabulated in table.\ref{table1} could be even lower.

\section{Summary and conclusions} 

To summarise, in the QGP motivated threshold model, we have analyzed the  mid-rapidity PHENIX data on the centrality dependence of $J/\psi$ suppression  in Au+Au  collisions. In the threshold model, in addition to the normal nuclear absorption, $J/\psi$'s are   anomalously suppressed, such that, if the local transverse density exceeds a threshold value $n_{J/\psi}$, all the $J/\psi$'s are absorbed. We have extended the threshold model to include the feedback from higher states $\chi$ and $\psi\prime$. The extended model has four parameters, 
the threshold density ($n_{J/\psi}$) for anomalous suppression of direct $J/\psi$, 
the threshold density ($n_\chi$) for anomalous suppression of the state $\chi$, the fraction $F$ of the states $\chi$ and
the $J/\psi$-nucleon absorption cross section ($\sigma_{abs}$). 

We  observed that the centrality dependence of $J/\psi$ suppression can not determine the fraction $F$ of the state $\chi$ and the nuclear absorption cross-section $\sigma_{abs}$ independently. In the threshold model they are inter-dependent. 
Anomalous suppression of the state $\chi$, to certain extent,  can mimic the  nuclear absorption of $J/\psi$ and $\chi$ and  
the PHENIX data can be fitted either with large $F$ and small $\sigma_{abs}$ or the vice-versa.
For $\sigma_{abs}$=0-2 mb, we have fitted the PHENIX data by varying $n_{J/\psi}$, $n_\chi$ and $F$. Data are well described for  $n_{J/\psi}=3.57\pm 0.17$ $fm^2$,  $n_\chi=0.32 \pm 0.32$ $fm^2$. 
The uncertainty in the parametric values could be reduced if the quality of the data is improved. 
The fraction $F$ depend on the value of $\sigma_{abs}$, $F=0.3\pm 0.1$, the upper and lower values corresponds to $\sigma_{abs}$=0 and 2 mb respectively.

Threshold model with parameters fixed from the Au+Au collisions is then used 
 predict for the centrality dependence of $J/\psi$ suppression in Pb+Pb collisions at LHC energy. At LHC energy
the model predicts that in central collisions $J/\psi$'s are
suppressed by a factor of 3-4. Suppression pattern is similar to that in Au+Au collisions at RHIC. In central collisions, possibly $J/\psi$'s are slightly less suppressed in Pb+Pb collisions at LHC energy than in Au+Au collisions at RHIC.  

We have also estimated the QGP formation time. On the basis of the empirical observation that in all the centrality ranges of Au+Au collisions, the square of the maximum transverse density 
times a factor 0.29, correctly reproduces the 
PHENIX measurements on Bjorken energy density times the formation time, we have 
argued that the threshold density gives the upper bound for the
QGP formation time. 
For critical temperature $T_c\approx=192$ MeV,
QGP formation time is estimated as $\tau\approx 0.09(0.07)$ fm.

To conclude, threshold model analysis of $J/\psi$ suppression in Au+Au collisions at RHIC, strongly suggest that in central and mid-central collisions  ($N_{part} \geq 150$) QGP is produced at sufficiently high temperature $T \geq 2.1 T_c$, to melt the directly produced $J/\psi$'s. The analysis also indicates that QGP is produced rather early in the collisions.


\end{document}